\documentclass{llncs} 

\usepackage{graphicx}
\usepackage{url}
\usepackage{amssymb,amsmath}
\usepackage{color}
\usepackage{float}
\usepackage{multicol}
\usepackage{multirow}

\begin{document}

\title{MT3S: Mobile Turkish Scene Text-to-Speech System for the Visually Impaired}
\author{Muhammet Ba\c{s}tan \and Hilal Kandemir \and B\"{u}\c{s}ra Cant\"{u}rk}
\institute{\email{mubastan@gmail.com, hkandemir0@gmail.com, busracanturk1@gmail.com}}

\maketitle

\begin{abstract}

Reading text is one of the essential needs of the visually impaired people.
We developed a mobile system that can read Turkish scene and book text, using a fast gradient-based multi-scale text detection algorithm for real-time operation and Tesseract OCR engine for character recognition. We evaluated the OCR accuracy and running time of our system on a new, publicly available mobile Turkish scene text dataset we constructed and also compared with state-of-the-art systems. Our system proved to be much faster, able to run on a mobile device, with OCR accuracy comparable to the state-of-the-art.

\end{abstract}

\section{Introduction}
\label{intro}

Reading text (scene text, book, newspaper, sign, banner, etc.) is one of the essential needs of the visually impaired people in daily life. With the advances in mobile devices with cameras, modest processing power and memory, it has become feasible to develop assistive mobile applications for the visually impaired. \textit{SayText}\footnote{http://www.docscannerapp.com/saytext} is a free IOS app that can read English book text for the visually impaired. \textit{Blind Navigator}\footnote{http://www.appcessible.org/apps/android/android-phone/blind-navigator} is a free Android app to provide a basic interface to an Android phone, like dialing numbers,  reading/composing messages by voice, GPS navigation, etc. \textit{LookTel}\footnote{http://www.looktel.com} is a commercial IOS app to recognize money and objects. There is no freely available assistive mobile app that can read both scene and book texts and supports multiple languages.

The goal of this work is to design and implement an assistive mobile text reading system that can read scene and book text and works on Android platforms (about 85\% of smart phones worldwide run Android operating system). Such a system has a number of requirements to be of practical use:
\begin{itemize}
 \item It should be able to run on a mobile platform with relatively limited resources (processing power, memory, battery).
 \item It should be fast (preferably real-time) and accurate.
 \item It should have a suitable user interface for the visually impaired.
\end{itemize}

\begin{figure}[h!]
	\centering
	\includegraphics[width=\textwidth]{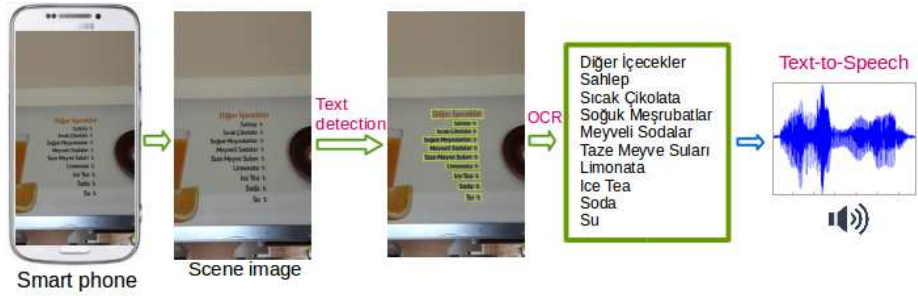}
	\caption{Mobile text detection, recognition and text-to-speech. As the scene is scanned with the device camera, the mobile app processes incoming scene images continuously to detect text regions. When text is detected, the user is notified with sound, text recognition is performed and resulting text is converted to speech.}
	\label{fig:system-flow}
\end{figure}

There are two types of usage scenarios for such a system. (1) The user has a text material (book, paper, etc.) at hand and wants to read aloud the text it contains. (2) The user wants to know if there is any informative scene text around her and read it aloud if there is. In the first case, the user has some rough idea where the text is, but not the exact location; in the second case, she does not even know if there is any text around. Hence, the mobile app should allow the user to scan the scene or document with the camera to locate any text and read it, and also guide her to get better images of the text for higher accuracy.

Optical character recognition (OCR) is the process of recognizing and converting image text into digital text form for further processing. For printed plain book text, OCR accuracy is very high, around 98-99\% for English~\cite{Tesseract-ICDAR07}. On the other hand, for natural scene text, it is still a challenging task and OCR accuracies are rather low due to many factors, including text transformations, font variations and size and adverse imaging conditions. OCR systems designed for printed text fail on natural scene text due to their text layout algorithms designed solely for printed text. Therefore, it is common practice to run a scene text detector and feed the detected locations to an OCR engine for character recognition. A recent trend is to use end-to-end text detection and recognition systems with the convolutional neural networks trained on large training data~\cite{wang-icpr12,text-survey-pami15,text-survey-fcs16}.

A typical mobile system to read text works as shown in Figure~\ref{fig:system-flow}. (1) Image acquired from the device camera is processed and text locations are determined (text detection). (2) Image and text locations are fed to an OCR engine and characters are recognized. (3) The recognized text is converted to speech using a text-to-speech (TTS) engine. In this pipeline, the speed and accuracy of text detection is critical for the success of the mobile system. As the user scans the scene or document with the device camera, the system should be able to accurately detect text regions in real-time and notify the user. If the system is slow, its practical use will be very limited.

In this paper, we describe our mobile Turkish scene text-to-speech (MT3S) system developed for the visually impaired users. It uses a fast and accurate text detection algorithm suitable for mobile platforms and off-the-shelf OCR and TTS engines for character recognition and speech synthesis. We constructed the first publicly available mobile Turkish scene text dataset and evaluated the text recognition performance of our system with comparisons with state-of-the-art systems. We also developed a proof-of-concept  Android app and demonstrated that our system works well in practice. Our system can easily be adapted to support other languages, especially those using derivatives of Latin or similar alphabets, with minor modifications.

\section{From Scene Text to Speech}
\label{sec:detection-ocr}

In this section, we describe our text detection and recognition system along with relevant work. We selected a pipeline approach, as in Figure~\ref{fig:system-flow}: first, detect text regions with a fast algorithm, then, recognize the text and convert it to speech. End-to-end text detection and recognition systems using convolutional neural networks (CNN) have been presented recently~\cite{wang-icpr12,text-read-ijcv2016} with state-of-the-art accuracy. However, currently, these systems have two major drawbacks: they need huge amounts of training data (and computational resources) and the CNN models are too large to be deployed on a mobile platform and run in real-time. Hence, further research is required to design smaller networks with high performance.

\subsection{Text Detection}
\label{sec:detection}

Many text detection algorithms have been proposed to date~\cite{text-survey-pami15,text-survey-fcs16}, using gradients/edges~\cite{phan2009,shivakumara2009,epshtein-swt-cvpr2010,canny-text-cvpr2016}, extremal regions~\cite{er-cvpr2012,portable-mva16}, CNNs~\cite{jaderberg-eccv14,text-cnn-tip2016}, etc. and it is still an active research area. Most of the methods utilize image gradients/edges or local regions with various post-processing operations to separate text from non-text regions.

The ER text detector~\cite{er-cvpr2012} is based on character detection using Extremal Regions (ERs).
The detection is a two stage sequential selection process. In the first stage, ERs are filtered with an AdaBoost classifier using simple to compute features, like aspect ratio, compactness, number of holes, etc.
In the second stage, computationally more expensive features with SVMs are used to improve the accuracy.
Finally, the surviving ERs are grouped into words to complete the text detection process. An OpenCV~\cite{OpenCV} implementation of ER text detector is publicly available\footnote{\url{https://github.com/opencv/opencv_contrib/tree/master/modules/text}}.

A relatively recent and quite accurate scene text detection system is SnooperText~\cite{snoopertext-cviu14}. It works in the following stages: segmentation of the image into foreground (text) and background, character filtering of foreground pixels with simple (height, aspect ratio) and more elaborate features (Fourier moments, pseudo-Zernike moments, polar encoding) fed to SVM classifiers, grouping of candidate characters into candidate text regions (words or text lines), filtering of candidate text regions with an SVM classifier using THOG descriptors and obtaining the final text regions to be recognized with an OCR engine.
SnooperText works quite well; it has publicly available Java source code\footnote{\url{http://www.dainf.ct.utfpr.edu.br/~rminetto/projects/snoopertext}}, but it is rather slow due to its complexity and also to Java implementation.

In our MT3S system, text detection speed is crucial, since, as the user scans the scene with the device camera, the system should be able to process the incoming images in real-time to decide if it contains text or not.
Therefore, slow text detection algorithms cannot be employed even if their accuracy is high. We should strike a balance between real-time operation and accuracy. Based on this consideration, we opted to devise an edge-based text detection algorithm using efficiently computable simple shape features to filter out non-text regions. The main idea is that text lines have regular horizontal structures with strong edges, mostly on plain background.

\begin{figure}[h!]
	\centering
	\includegraphics[width=\textwidth]{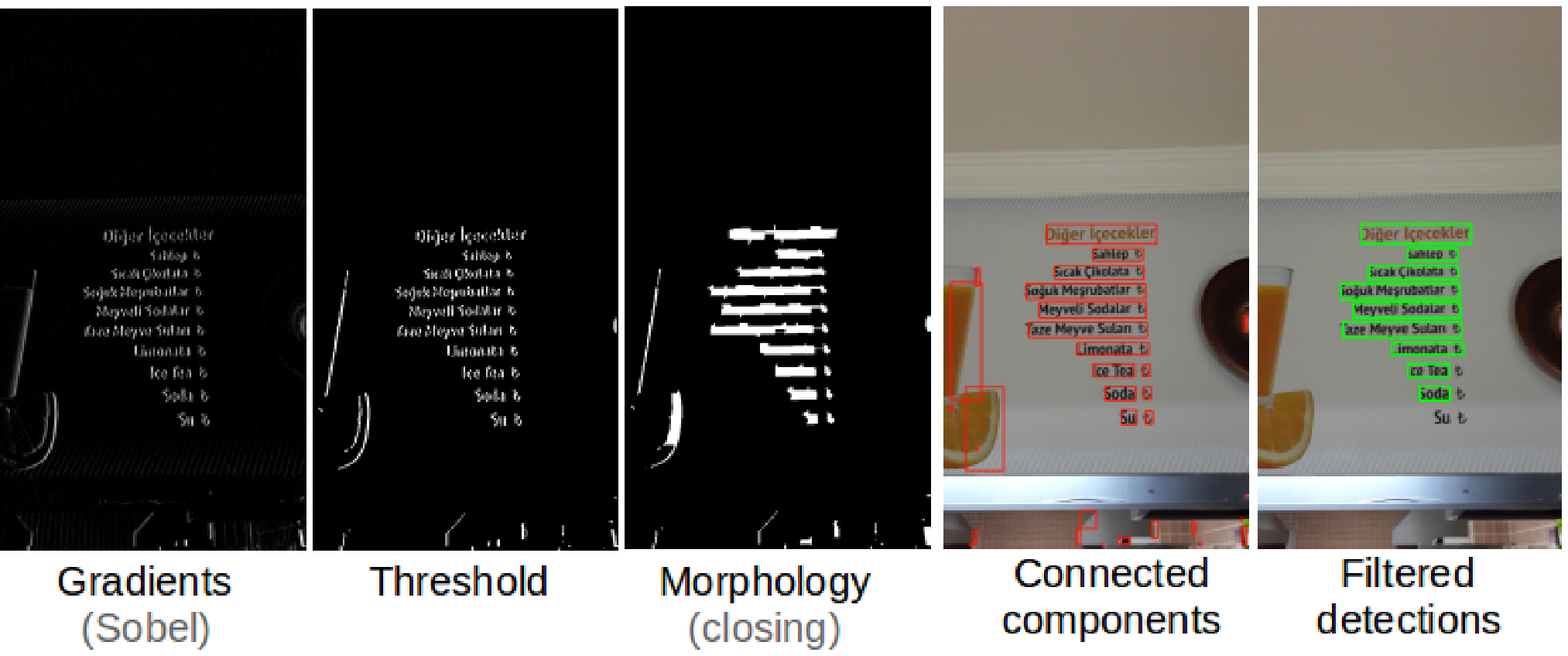}
	\includegraphics[width=\textwidth]{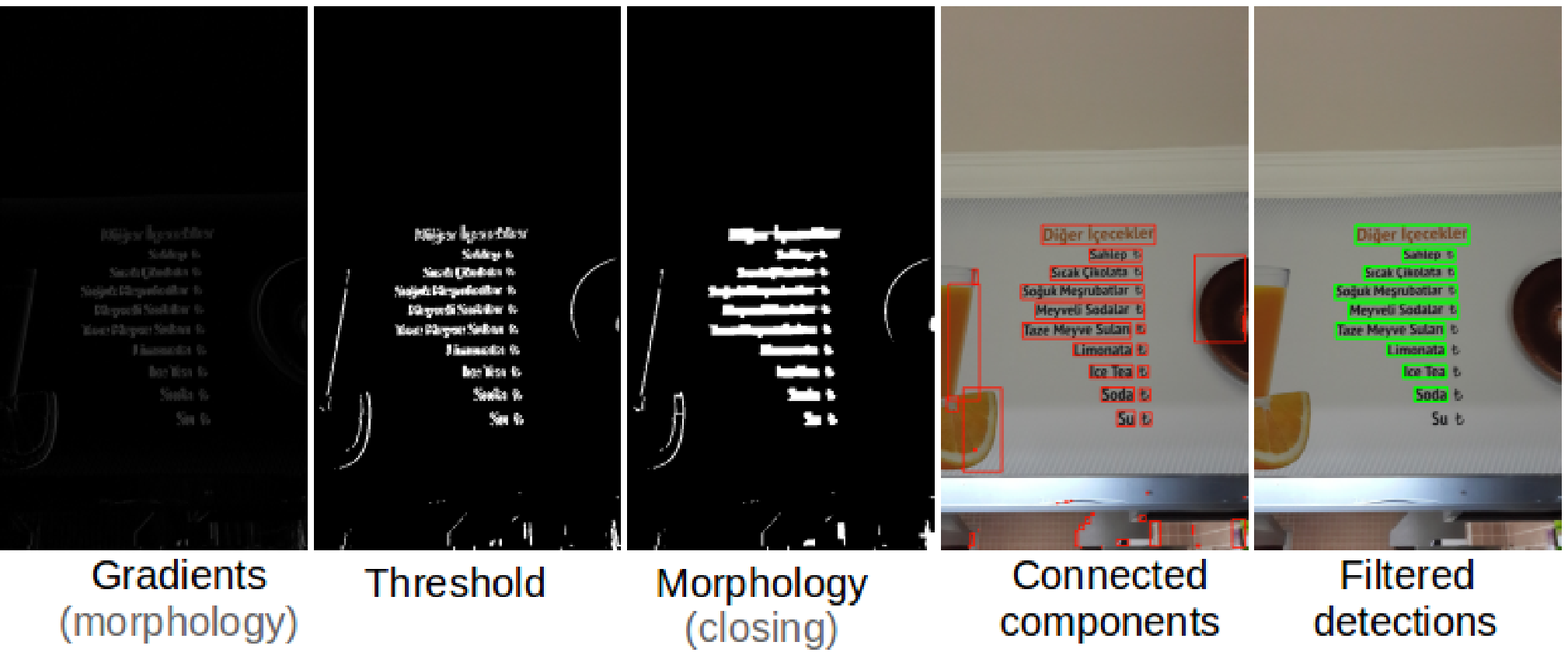}
	\includegraphics[width=0.8\textwidth]{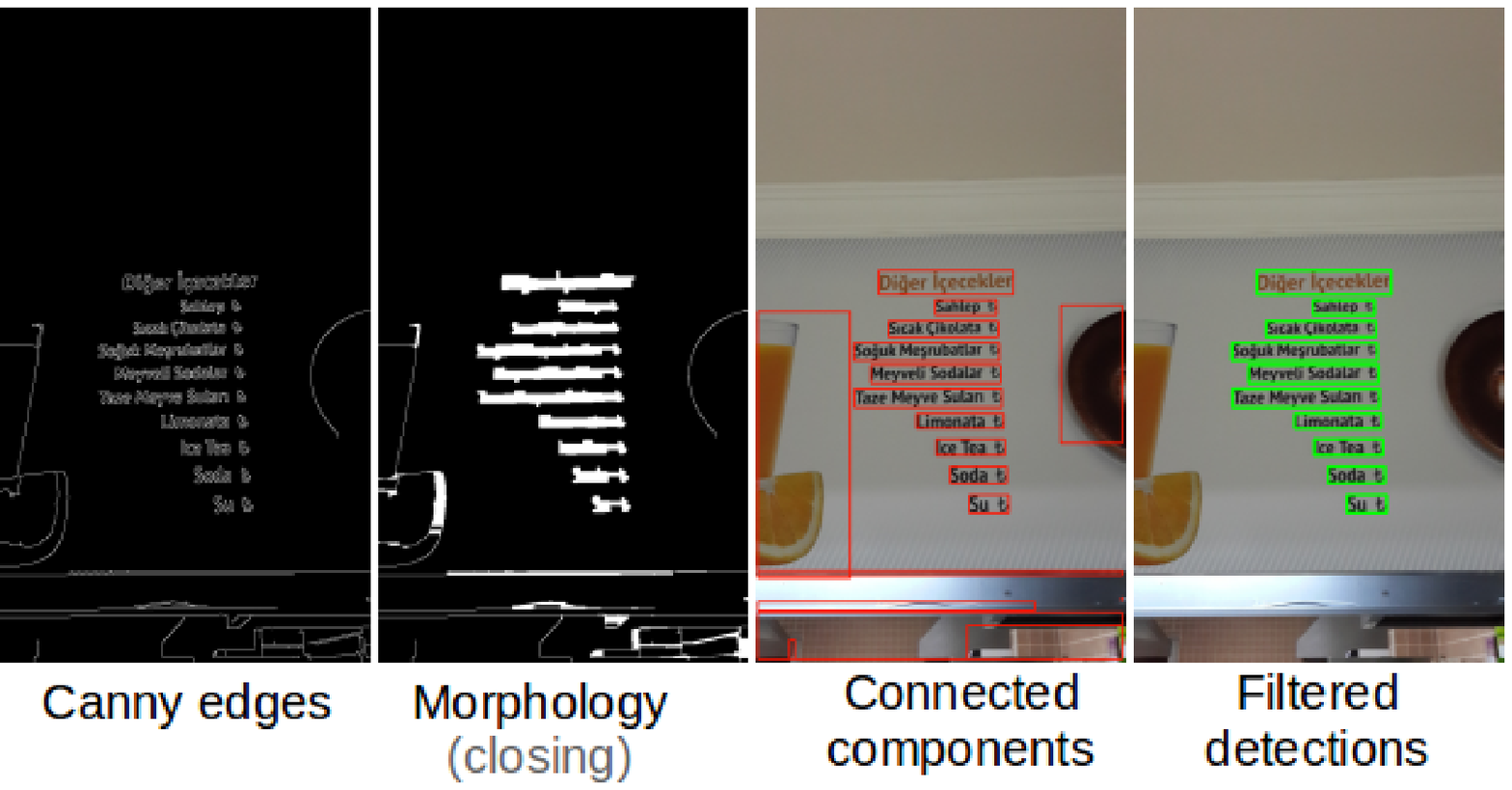}
	\caption{Text detection procedure using Sobel, morphological gradients and Canny edge detection. The final detections are shown on the right with green bounding boxes.}
	\label{fig:text-detection}
\end{figure}

The major processing steps of our edge-based text detection algorithm is shown in Figure~\ref{fig:text-detection}: (1) computing gradients/edges, (2) finding text regions with morphology, (3) labeling connected components and text/non-text filtering.

\subsubsection{Gradient/Edge Computation}

Text regions have strong gradients, mainly in the horizontal direction ($G_x$). Therefore, we compute $G_x$ gradients using Sobel operator with an aperture size of 3 or with a morphological operator with a kernel size of $3\times1$. For RGB images, the gradients on R, G, B channels are computed separately and their maximum is taken, $G_x = max(G_{Rx}, G_{Gx}, G_{Bx})$. Then, the gradient image is thresholded using the OTSU algorithm to obtain a binary image (Figure~\ref{fig:text-detection}). The foreground pixels correspond to high-gradient edge pixels, which may belong to characters.

In addition to the thresholded gradients, we also experimented with Canny edges, with aperture size 3 and low, high thresholds 50, 200.

\subsubsection{Finding Text with Morphology}

We apply morphological closing operation on the edge image to merge edge pixels in the horizontal direction, since text lines are mostly horizontal (hence, we aim to find only horizontal or near horizontal text lines). We use a long rectangular structuring element of size $17\times5$ on Sobel and Canny edges, and $11\times5$ on morphological edges. Thresholded morphological edges are thicker, therefore, a smaller structural element is sufficient and also faster. The sizes are determined experimentally. After this operation, horizontal text lines stand out as horizontal rectangular blocks in the output image, as shown in Figure~\ref{fig:text-detection}. However, there are also other foreground non-text regions, which need to be filtered out.

Text sizes are variable. With a fixed size structural element, it is only possible to detect texts of certain sizes. Therefore, a multi-scale strategy is needed to detect texts at various sizes. We address this issue in Section~\ref{sec:multi-scale} below.

\subsubsection{Connected Components and Text/Non-Text Filtering}
As shown in Figure~\ref{fig:text-detection}, after morphological closing, there are many non-text foreground regions remaining. To filter out non-text regions, we first label the foreground connected components and compute simple shape features from each component to decide if it is text or not. For each component, we apply the tests from simple to slightly complex, in the following order for efficiency reasons.

\begin{itemize}
 \item \textit{Area.} If the component is too small (area smaller than 0.001 of the image area), it is discarded.
 \item \textit{Height.} If the height of the component is less than 17 pixels or larger than 0.25 of the minimum image side, $min(width,height)$, it is discarded.
 \item \textit{Aspect ratio.} If aspect ratio ($width/height$) of the component's minimum bounding rectangle (MBR)  is less than $1.3$, it is discarded. Text lines mostly have high aspect ratio.
 \item \textit{Extent.} Extent is defined as the ratio of the number of foreground pixels in the component to the area of its bounding rectangle. Extent for a text line should ideally be close to $1.0$, since text lines appear as filled rectangular regions after morphological closing. We used an extent threshold of $0.4$, and discarded components with lower extent values. We computed two extent values, one on the morphologically closed edge image and one on the thresholded edge image (with a smaller extent threshold value $0.3\times0.4$). Finally, the extent value of the components is weighted with the logarithm of its aspect ratio ($log(ar)\times extent$) to promote components with high aspect ratio but have small extent value due to slight text rotation and hence a large MBR area.
\end{itemize}

If a connected component passes all these tests, it qualifies as a text region (last column in Figure~\ref{fig:text-detection}). The tests require very simple shape features, easily computable with a labeled connected component image. It is also easy to implement using the available OpenCV functions. We could also use a cascaded classifier with these simple features to learn the best thresholds on a specific dataset, but sufficiently large labeled training data is required. With a simple classifier, the detection speed should be close to the above algorithm.

A text region is represented with the minimum bounding rectangle  of its connected component. We observed that, the bounding rectangles are slightly smaller than the actual text region and also missing the left part of the initial letter of the text region. Therefore, we enlarge the detected bounding boxes, first by $0.1$ of text height in $-x$ direction (to left), and then by $0.05$ of text height in all directions. This is also helpful, later for correctly recognizing some of the non-ASCII Turkish characters \c{C}, \u{G}, \.{I}, \"{O}, \c{S}, \"{U}, which may be overflowing the MBR.

\subsection{Multi-Scale Text Detection}
\label{sec:multi-scale}

\begin{figure}[h!]
	\centering
	\includegraphics[width=0.8\textwidth]{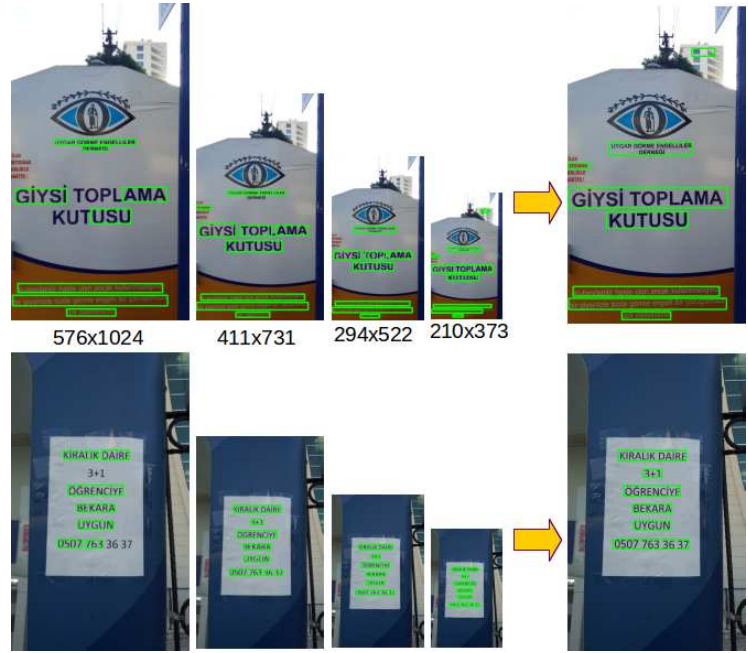}
	\caption{Multi-scale text detection. The original image is downsized by 1.4 at each step until one side becomes less than 200 pixels. Detections at each scale are combined to obtain the final detections, shown on the right.}
	\label{fig:ms-text-detection}
\end{figure}

As discussed above, using a fixed size structuring element, it is only possible to detect texts of a certain size, larger text regions will go undetected. The text detection method described above is single scale, a multi-scale strategy is needed to detect texts of various sizes. Figure~\ref{fig:ms-text-detection}, left image, shows the result of text detection on the original size image; some of the large text regions could not be detected. As a solution, we apply multi-scale text detection: repeatedly downsize the image $1.4$ times until one side of the image becomes less than 200 pixels and apply single scale text detection at each scale. At each scale, we adjust the area and height thresholds accordingly. We also use smaller structuring elements, $15\times3$ for Sobel gradients and Canny edges, $9\times3$ for morphological gradients (fixed size at all scales).

Finally, detections in all scales are combined to obtain the final detections. If a text region is inside another one, it is discarded. If two text regions have high area overlap ($>0.80$), smaller one is discarded. Figure~\ref{fig:ms-text-detection} shows two examples of multi-scale text detection, in which some text regions not detected in the original image could be detected in the downsized images.

\subsection{Text Recognition and Text-to-Speech}

The ultimate goal is to recognize the text (OCR) in an image so that it can be converted to speech.
After localizing the text regions, their MBRs are fed to the Tesseract OCR engine for recognition (currently, we do not apply any correction on rotated text regions). Then, Tesseract processes only the given text regions in the image. The page segmentation mode is set to ``single block'' (PSM\_SINGLE\_BLOCK), in case some of the detected regions contain multiple text lines. There are other page segmentation options, like ``single line'' (PSM\_SINGLE\_LINE), ``single word'' (PSM\_SINGLE\_WORD), etc.

Finally, the recognized texts are sorted in reading order, from top to bottom and from left to right, and given to a text-to-speech engine for speech synthesis.

\section{Dataset and Annotation}
\label{sec:dataset}

We constructed the first Mobile Turkish Scene Text (MTST 200) dataset, consisting of 200 indoor and outdoor Turkish scene text images. The labeled dataset is publicly available for research purposes\footnote{\url{https://github.com/mubastan/mtst200}}.
The images were collected with mobile phones and downsized to $576\times1024$ (portrait) or $1024\times576$ (landscape) pixels. The text lines are horizontal or near horizontal, some with slight in-- and out-of-plane rotations.

For OCR performance evaluation, we labeled the MTST 200 dataset with the Scene Text Annotation Tool (STAT) we developed to annotate scene text images with bounding boxes or bitmaps. STAT was written in Python with PyQT4 and made publicly available for research purposes \footnote{\url{https://github.com/mubastan/stat}}. It supports UTF-8 labels so that it can be used to label text in languages that need UTF encoding, like Turkish and German.

\begin{figure}[h!]
	\centering
	\includegraphics[width=0.9\textwidth]{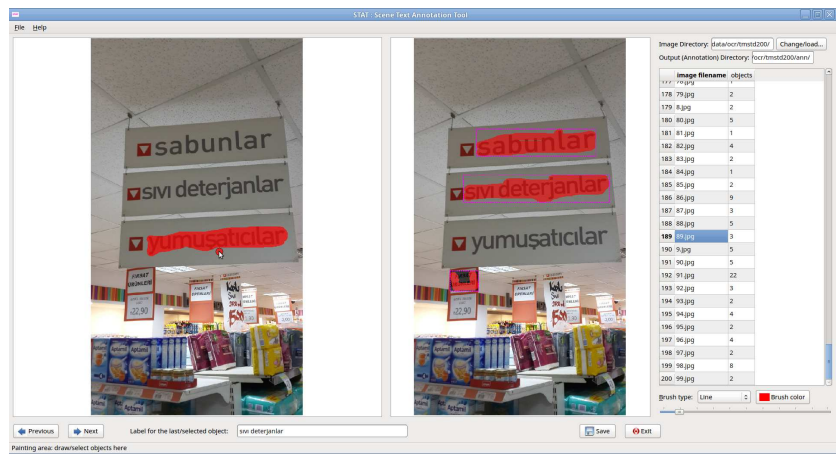}
	\caption{STAT: Scene Text Annotation Tool screen shot.}
	\label{fig:stat}
\end{figure}

\section{Experiments}
\label{sec:experiments}

\subsection{Setup and Evaluation}

The goal of our system is to detect and recognize texts in images coming from a mobile device camera and convert the texts to speech. Therefore, text recognition accuracy and operation speed are the most important performance metrics for evaluation. We performed experiments on our MTST 200 dataset and measured these performance metrics.
To this end, we implemented our system using the OpenCV (v3.1)~\cite{OpenCV} and Tesseract (v3.0.4)~\cite{Tesseract} libraries. We compared our system with existing systems in Section~\ref{sec:comparison} below.

We first run the text detector on each image to find text locations, on which we run Tesseract to produce text outputs for each text location. Then, we sorted the text outputs for each image in reading order (top to bottom, left to right) and saved to a text file. Finally, we used the ``ISRI OCR Evaluation Tool''~\cite{ISRI} to compare the OCR output files with the ground truth labels for each image. The comparison is case-sensitive (though, case-sensitivity is not crucial in our case). The overall OCR accuracy is obtained over the whole dataset by combining the ISRI evaluation results from all the files.

The OCR accuracy for a text with $n$ characters is computed as $\frac{n-e}{n}$, where $e$ is the number of erroneous characters, which is the number of edit operations (insertions, deletions, substitutions) required to correct the text. For some image texts, $e$ turns out to be greater than $n$ (due to false positives) and the result is negative, which means text recognition failed. We set $e=n$ for such images (zero accuracy), so that failure on these images does not affect the accuracy on the others.

\subsection{OCR Results}

\begin{table}
  \centering
  \caption{OCR accuracy of MT3S (this work) on MTST 200 dataset with single and multi-scale methods on grayscale and RGB gradients/edges.}
  \label{table:ocr-result}
  \resizebox{0.8\textwidth}{!}{
    \begin{tabular}{lcccccc}
      \hline	
	  & Sobel    & Sobel    & Morphology & Morphology        & Canny    & Canny   \\
	  & (single) & (multi)  & (single)   & (multi)           & (single) & (multi) \\
      \hline
    Gray  & 42.45    & 53.74    &   48.05    &  53.98            &  44.23   & 48.59  \\
    RGB   & 44.15    & 51.88    &   48.51    &  \textbf{55.61}   &    -     &  -     \\
    \end{tabular}
	}
\end{table}

\begin{table}
  \centering
  \caption{Average running times per image (milliseconds) of text detection methods in MT3S on MTST 200 dataset.}
  \label{table:runtime}
  \resizebox{0.8\textwidth}{!}{
    \begin{tabular}{lcccccc}
      \hline	
	  & Sobel    & Sobel   & Morphology & Morphology & Canny    & Canny   \\
	  & (single) & (multi) & (single)   & (multi)    & (single) & (multi) \\
      \hline
    Gray  & 15.81    &  32.57  & 9.17      & 23.51       &   9.52   &  22.36 \\
    RGB   & 30.65    &  69.03  & 12.65     & 31.83       &   -      &  - \\
    \end{tabular}
	}
\end{table}

Table~\ref{table:ocr-result} shows OCR accuracies of our system on the MTST 200 dataset, with various gradient/edge computation methods using grayscale and RGB color images. The highest accuracy is obtained with morphological gradients on RGB color images with multi-scale detection. Using horizontal $G_x$ gradient edges turns out to be better than using Canny edges. Multi-scale detection improves detection and hence OCR accuracy significantly. The affect of multi-scale detection depends on the font size distribution of the dataset. If the dataset contained only fixed size small text, then multi-scale detection would not have improved the accuracy. However, in practice, text sizes vary considerably, especially in a mobile scene text-to-speech system.

Table~\ref{table:runtime} shows average running times of text detection methods on MTST 200 dataset, measured on an Intel CORE i7 laptop with 32GB memory. OpenCV implementation takes advantage of multi-core architectures for parallel processing, with significant speedup. According to the table, morphological and Canny edges are faster than Sobel edges. Considering the OCR accuracy and running times together, morphological edge detection should be preferred, since it is both faster and more accurate.

\subsection{Comparison}
\label{sec:comparison}

In this section, we compare the OCR accuracy and running time performance of our MT3S system with two state-of-the art text detection systems, ER detector~\cite{er-cvpr2012} and SnooperText~\cite{snoopertext-cviu14}. We selected these two systems for comparison, because (1) their source codes are publicly available, (2) they reported high performance in their papers~\cite{er-cvpr2012,snoopertext-cviu14}.

ER detector has an OpenCV implementation\footnote{\url{https://github.com/opencv/opencv_contrib/tree/master/modules/text}}. SnooperText has a Java implementation by its authors \footnote{\url{http://www.dainf.ct.utfpr.edu.br/~rminetto/projects/snoopertext}}. Both use RGB color images as input.

The OpenCV implementation of ER detector produces multiple overlapping bounding rectangles for text regions; we filtered out the overlapping detections by using the same algorithm as our system uses to combine multi-scale detections (smaller text regions are discarded in case of high overlap).

The Java implementation of SnooperText is rather slow with default parameters. We decreased the number of pyramid levels from 10 to 5, and increased the minimum text height from 13 to 17, so that it is faster and also work in similar conditions as MT3S. The rest of the parameters were not changed.

We performed two more experiments on the dataset, corresponding to two extreme cases, one without any text detections and one with manual annotations (close to perfect text detection). (1) We run Tesseract OCR engine directly on the raw images, without any text detection; this can be considered to be a baseline. (2) We run Tesseract on the ground truth (manual) text annotations, to see what would be the OCR accuracy of Tesseract if the detections were perfect, assuming the manual annotations are close to perfect (though manual annotations are also never perfect).

\begin{table}
  \centering
  \caption{Comparison of OCR accuracies of MT3S (this work) with existing systems on MTST 200 dataset.}
  \label{table:ocr-result-comp}
  \resizebox{0.8\textwidth}{!}{
    \begin{tabular}{ccccc}
      \hline	
	  Manual    &  Tesseract  & SnooperText~\cite{snoopertext-cviu14}   &   ER~\cite{er-cvpr2012}   &   MT3S (this work) \\
      \hline
	  72.14     & 19.23        &   60.97       &  49.77  &    55.61    \\
    \end{tabular}
	}
\end{table}

\begin{table}
  \centering
  \caption{Comparison of average text detection times per image (milliseconds) on MTST 200 dataset.}
  \label{table:runtime-comp}
  \resizebox{0.55\textwidth}{!}{
    \begin{tabular}{ccc}
      \hline	
	 SnooperText~\cite{snoopertext-cviu14}   &   ER~\cite{er-cvpr2012}   &   MT3S (this work) \\
      \hline
	   7240   &  1219  &    31.83    \\
    \end{tabular}
	}
\end{table}

Tables~\ref{table:ocr-result-comp} and \ref{table:runtime-comp} show the OCR accuracies and average running times of compared methods on MTST 200 dataset. Based on the results in the tables:
\begin{itemize}
 
 \item Running Tesseract directly on the raw images, without applying text detection, results in a very poor average OCR accuracy of 19.23\%. Tesseract works very well on text images with plain background, but fails with cluttered backgrounds, since it is designed for plain document images.
 
 \item With manual annotations, OCR accuracy is 72.14\%, which can be assumed to be the upper bound on this dataset. This shows roughly the OCR accuracy limit of Tesseract on the dataset.
 
 \item SnooperText works quite well with 60.97\% OCR accuracy, but it is rather slow due to its complexity and Java implementation. It is probably possible to speed it up with parallel implementations, e.g., with OpenCV, but it is unlikely to be fast enough to run in real-time.
 
 \item The ER detector is both slow and did not perform well on the dataset.
 
 \item Our system MT3S works well, close to SnooperText with 55.61\% OCR accuracy and very fast, with 31.83 milliseconds per image (227 times faster than SnooperText, 38 times faster than ER detector), it can run in real-time. This is the best result of MT3S using multi-scale detection with morphological RGB edges.

\end{itemize}

\begin{figure}[h!]
	\centering
	\includegraphics[width=0.8\textwidth]{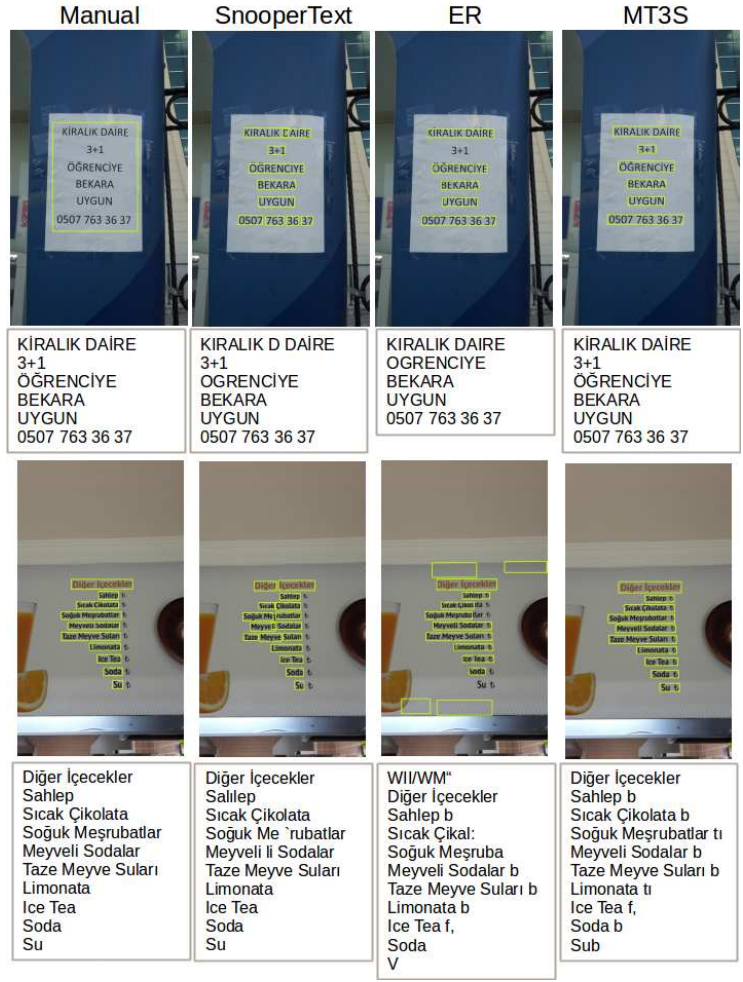}
	\caption{Sample text detection and OCR results from MTST 200 dataset. From left to right, manual annotations, SnooperText~\cite{snoopertext-cviu14}, ER detector~\cite{er-cvpr2012}, MT3S (this work), respectively. }
	\label{fig:samples1}
\end{figure}

\begin{figure}[h!]
	\centering
	\includegraphics[width=0.8\textwidth]{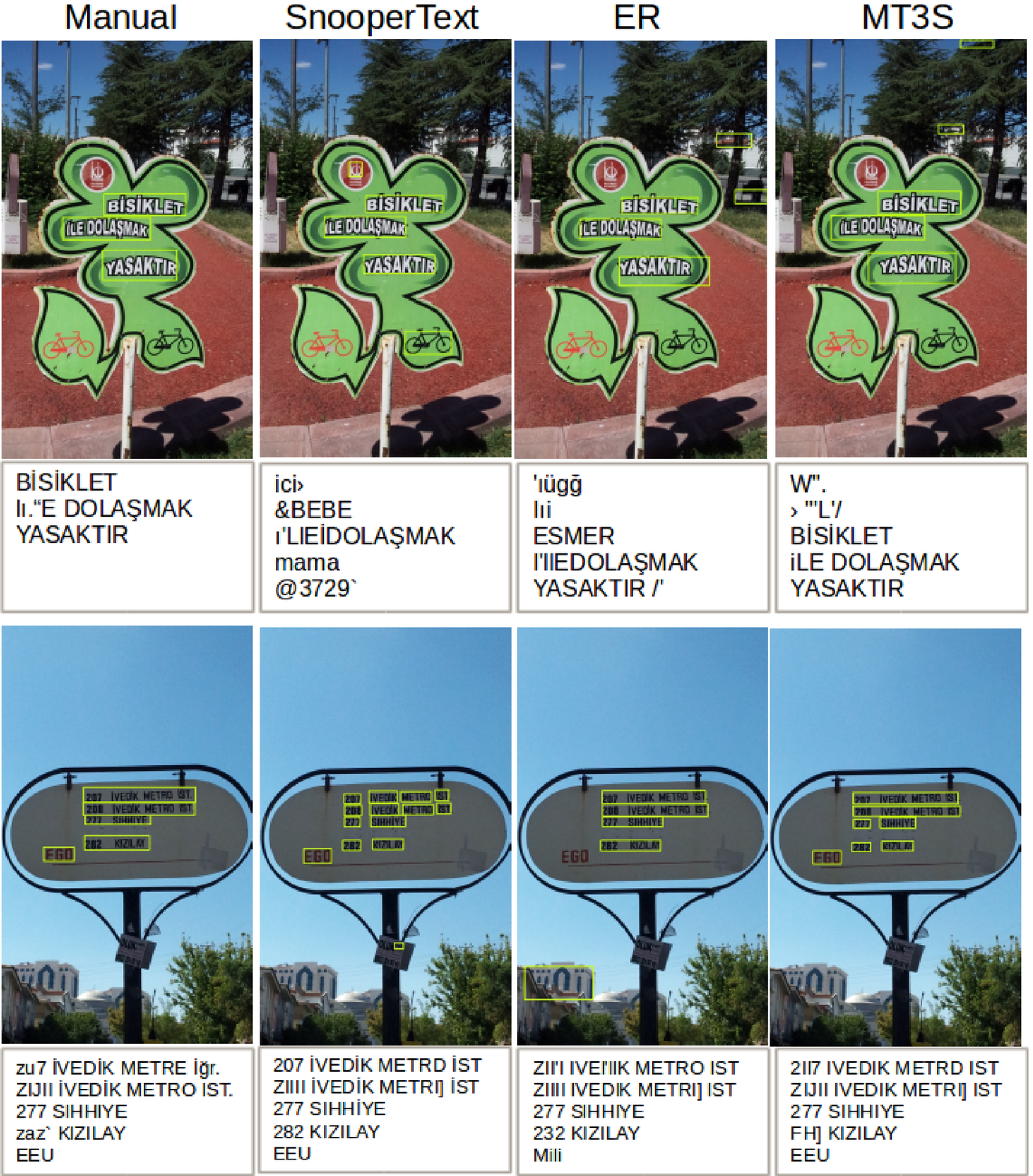}
	\caption{Sample text detection and OCR results from MTST 200 dataset. From left to right, manual annotations, SnooperText~\cite{snoopertext-cviu14}, ER detector~\cite{er-cvpr2012}, MT3S (this work), respectively.}
	\label{fig:samples2}
\end{figure}

Figures~\ref{fig:samples1} and \ref{fig:samples2} show sample text detection and OCR results from the MTST 200 dataset. On images with plain backgrounds, all detection methods work well and Tesseract can recognize the characters well, but on cluttered images, even small differences in detected text regions seem to affect the OCR accuracy of Tesseract.

\subsection{How to Improve Text Detection and OCR Accuracy}
\label{sec:improve}

As discussed in the previous section, even with perfect text detection, the OCR accuracy is 72.14\% on MTST 200 dataset (Table~\ref{table:ocr-result-comp}). On the other hand, OCR accuracy of Tesseract on English document images is about 98-99\%~\cite{Tesseract-ICDAR07}. Lower OCR accuracy is mainly due to text rotations, cluttered backgrounds and adverse imaging conditions. One way to improve accuracy would be to correct rotated text regions before feeding to Tesseract; this is not done our experiments. OCR on scene text images seems to be the major bottleneck; it is currently an active research area to improve OCR accuracy on challenging scene text images with varying fonts, font sizes, image clutter and adverse imaging conditions~\cite{text-survey-fcs16}.

The OCR accuracy of Tesseract on a scanned Turkish document images dataset was reported to be 86.41\% in~\cite{karasu-siu15}; this is significantly lower than the accuracy on English datasets. One possible explanation may be the quality of trained Turkish model is not as good as the English model, therefore, it may be possible to get higher accuracies with a better model.

Compared to OCR accuracy on manual annotations, performance loss due to inaccurate text detection in our system is about 16\%. We observed that the major weakness of our text detection system is that when text regions are very close to non-text edges, text and non-text regions are merged during morphological closing into one connected component and discarded (missed text regions). One possible solution would be to compute projection profile in such cases and segment out  text from non-text regions, but this will increase the computational cost.

\section{Android Application}
\label{sec:app}

\begin{figure}[h!]
	\centering
	\includegraphics[width=0.8\textwidth]{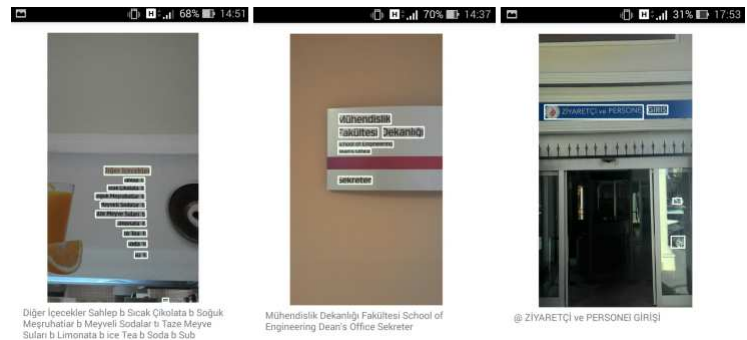}
	\caption{Screenshots from the text-to-speech screen of the prototype Android application (developed by the second author).}
	\label{fig:app-sshot}
\end{figure}

As a proof of concept, we have developed an Android app, called \textit{Mobile Reader}, with a suitable user interface for the visually impaired. The app accepts tactile inputs and produces speech outputs. There are three main screens, corresponding to three modes of operation: idle, scene scanning and text-to-speech. The user can switch between the modes with left-right swipes. The app opens in idle mode and waits for user input. In scene scanning mode, the user scans the scene with the device camera; when the app detects text in the scene, it notifies the user with sound and transitions to the text-to-speech mode. The detected text is recognized and converted to speech. The user can listen to the speech again or return back to the previous operation modes. In scanning mode, the user is guided to get better images of the text, when the text is on image borders. Figure~\ref{fig:app-sshot} shows screenshots from the prototype application's text-to-speech screen, where detected and recognized text are shown on the screen for test purposes (in actual use, no visual output is needed).

We implemented the Android app using the Android ports of OpenCV and Tesseract\footnote{\url{https://github.com/rmtheis/tess-two}}. The current version uses multi-scale detection with Sobel edges on grayscale images. For text-to-speech, we used Android text-to-speech (TTS) library, which supports Turkish with satisfactory speech quality. There are other high quality TTS libraries that support Turkish, like MaryTTS\footnote{\url{http://mary.dfki.de}}, but their Android ports are not available yet. The app can read both book and scene text.

The current unoptimized Android implementation can run text detection at 0.24 seconds per image (on ASUS ZENPHONE6, 1.6GHz processor, 2GB memory), i.e., about 4-5 frames per second. This is not real-time, but turned out to be satisfactory in our tests to slowly scan the scene for text detection. The running speed can be improved using morphological edges and more optimized implementation and detection parameters.

\section{Conclusions}
\label{sec:conclusion}

We designed and implemented a mobile scene text-to-speech for the visually impaired and tested it on a Turkish scene text dataset. The system works well for both scene and book texts and can easily be adapted to other languages, provided they are supported by the Tesseract OCR and Android TTS engines.
The accuracy of the system can be improved by improving the text detection accuracy, training a better language model for Tesseract, and post-processing the recognized text to correct errors.

\bibliographystyle{spmpsci}
\bibliography{References}

\end{document}